\input amstex
 \documentstyle{amsppt}
 \nopagenumbers \magnification=1100
 \centerline{\bf  On the Determination of Moving Boundaries}
 \centerline{\bf  for Hyperbolic Equations} \vskip.1in \centerline{Gregory Eskin and James
Ralston}

\centerline{Department of Mathematics, UCLA}
\vskip.2in \vskip.2in \noindent{\bf Abstract.} We
consider wave equations in domains with
time-dependent boundaries (moving obstacles)
contained in a fixed cylinder for all time. We
give sufficient conditions for the determination
of the moving boundary from the Cauchy data on
part of the boundary of the cylinder. We also
study the related problem of accessibility of the
moving boundary by time-like curves from the
boundary of the cylinder. \vskip.2in

\centerline{\bf \S 1. Introduction} \vskip.1in In
this article we study the possibility of
determining a moving boundary from Cauchy data on
a stationary boundary. The setting of this
problem is as follows. Let $Q$ be a connected
domain in $\Bbb R_x^n\times \Bbb R_t$ with smooth
boundary $\partial Q$. We assume that the
complement of $Q$ is contained in the cylinder
$C=\{(x,t): |x|<\rho,\ t\in \Bbb R\}$, and that
$\partial Q$ is moving at speed uniformly less
than 1, i.e. if $\nu=(\nu_x,\nu_t)$ is a normal
vector to $\partial Q$, then $|\nu_t|\leq r|\nu_x
|$ for a fixed $r<1$. This condition means that
$\partial Q$ is \lq\lq time-like". These
assumptions imply that the sets $\Omega_t=\{x:\
(x,t)\in Q\}$ are diffeomorphic and connected. We
think of the complement $\Omega_t^c$ as an
impermeable obstacle which is smoothly deformed
to $\Omega_{t^\prime}^c$ as time goes from $t$ to
$t^\prime$.

The assumption that $\partial Q$ is time-like
implies that the  following boundary value
problems are well-posed for $f\in
C_c^\infty(\partial C)$: the forward problem
$$u_{tt}-\Delta u=0 \hbox{ in } Q\cap C,\ u=0 \hbox{ on }
\partial Q,\ u=f \hbox{ on }
\partial C,\hbox{ and }u=0 \hbox{ when }t<<0,\eqno{(1)}$$
and the  backward problem $$u_{tt}-\Delta u=0
\hbox{ in } Q\cap C,\ u=0 \hbox{ on }
\partial Q,\ u=f \hbox{ on }
\partial C,\hbox{ and }u=0 \hbox{ when }t>>0.\eqno{(1^\prime)}$$
Letting $u^f$ denote the solution to the forward
problem with boundary data $f$, we have the set
of Cauchy data on $\partial C$
$$\Cal K(Q) =\{ (f,{\partial u^f\over \partial \nu})\vert_{\partial C}: f\in
C^\infty_c(\partial C)\}.$$ With these
definitions  one can ask the question: does $\Cal
K(Q)$ determine $Q$?

 Surprisingly, even when the motion is periodic, i.e. when $Q$ is
invariant under the mapping $t\to t+1$, the
answer is no. This was discovered  by Stefanov in
[St]. So the general problem is to characterize
the $Q$ which are determined by $\Cal K(Q)$. This
still appears quite difficult. Here we will
restrict ourselves to a discussion of sufficient
conditions for the Cauchy data to determine $Q$.

A  condition more easily verified than $\Cal
K(Q)$ determines $Q$ is that all points of
$\partial Q$ are accessible both forward and
backward in time. We define $(x_0,t_0)\in
\partial Q$ to be accessible (forward) if there
is a piecewise differentiable curve $(x(t),t)$ in
$Q$ with $|\dot x(t)|\leq 1$ such that
$x(t_0)=x_0$ and $(x(t_1),t_1)\in C$ for some
$t_1<t_0$. Accessible backward is defined the
same way with $t_1>t_0$. The failure of
accessibility is essential in Stefanov's example.
In \S 2 we discuss accessibility in two space
dimensions. We also show that all points will be
accessible if one can parameterize $\partial Q$
in the following way: suppose that $(x,t)\in
\partial Q$ is equivalent to $x=\phi(y,t)$, where $\phi(\cdot,t)$ is a smooth diffeomorphism
of $\partial\Omega_0$ into $\Bbb R^n$ satisfying
$\phi(y,t)=\phi(y,t+1)$ and $|\phi_t(y,t)|<1$. In
other words $\partial Q$ is moving periodically,
and has a periodic parametrization with time
derivative having norm less than one (see
Proposition 2.2). Note that the assumption that
$\partial Q$ is moving at speed less than one
only implies that the normal component of the
time derivative of $\phi$ has norm less than one.
In \S 3 we present a simplified version of
Stefanov's example.

 An implicit sufficient condition  for $\Cal K(Q)$ to
determine $Q$ is that there is a diffeomorphism,
$\Phi(y,t)=(\Psi(y,t),t)$ such that $\Psi(y,t)=y$
on $\partial C$, $\Psi(\cdot,T)$ maps $Q\cap
C\cap \{t=0\}$ onto $Q\cap C\cap \{t=T\}$ for all
$T$, and satisfies the following (surprisingly
restrictive) conditions: $|\Psi_t(y,t)|<1$ and $
||\Psi_y(y,t)||\leq M<\infty$ for all $(y,t)$.
Here $\Psi_y$ is the Jacobian matrix. We give the
proof of this result in \S 4 for more general
hyperbolic operators of the form
$$\partial^2_t u-2a(t,x)\cdot \nabla_x
 \partial_tu -\nabla_x\cdot A(t,x)\nabla_x u
-L_1(t,x,\partial_t,\partial_x)u,\eqno{(2)}$$
where $A$ is positive definite and $L_1$ is a
differential operator of order one. We also
discuss four situations where one can construct
$\Psi$ with the properties mentioned above.

There is substantial literature on wave equations
in domains with moving boundaries. In particular,
the fundamental existence and uniqueness results
were established by Ikawa [I], and  Cooper and
Strauss developed  scattering theory for these
equations and solved the inverse problem for
moving convex boundaries in a series of papers
(see [CS1-3]). Their proof also shows that $\Cal
K(Q)$ determines $Q$ when $\partial \Omega_t$ is
convex for all t
 \vskip.2in
\S 2. {\bf Periodic Obstacles in Two Space
Dimensions: Accessibility} \vskip.2in In this
section we consider domains $Q\subset \Bbb
R^2_x\times \Bbb R_t$ with boundary $\partial Q$
given by $(x(\sigma,t),t),\ \sigma\in S^1$. We
assume that $x(\sigma,t)$ is smooth,
non-degenerate: $|x_\sigma|>0$, and periodic:
$x(\sigma,t+T)=x(\sigma,t)$. The boundary of $Q$
will be time-like for $u_{tt}-\Delta$ precisely
when the normal component of $x_t$ is strictly
less than one, i.e. when $|x_t-(x_t\cdot
x_\sigma)|x_\sigma|^{-2}x_\sigma|<1$. This
condition can be stated in several equivalent
forms. One that is useful is
$$|x_\sigma|^2|x_t|^2<(x_t\cdot x_\sigma)^2+|x_\sigma|^2.\eqno{(3)}$$
Instead of taking $\sigma\in S^1$, it is
convenient to use the equivalent formulation
$\sigma\in \Bbb R$ with
$x(\sigma+1,t)=x(\sigma,t)$. Note that the
periodicity in both $t$ and $\sigma$ implies that
the non-degeneracy and time-like conditions hold
uniformly: $|x_\sigma|^2\geq \delta>0$ and
$|x_\sigma|^2+(x_t\cdot
x_\sigma)^2-|x_\sigma|^2|x_t|^2\geq \delta>0$ for
all $(\sigma,t)$ for some $\delta>0$.

Suppose that $x(\sigma(t),t)$ is a null geodesic
(light curve) in the Minkowski metric
$(dt)^2-(dx_1)^2-(dx_2)^2$ restricted to
$\partial Q$. Then
$$0=1-|x_t|^2-2(x_\sigma\cdot x_t)\sigma^\prime
-|x_\sigma|^2(\sigma^\prime)^2,$$ and we have two
differential equations for possible $\sigma(t)$'s
$${d\sigma_\pm\over dt}=\Lambda_\pm(\sigma,t),\hbox{ where }$$
$$\Lambda_\pm={-x_\sigma\cdot x_t\pm \sqrt{(x_\sigma\cdot
x_t)^2+(1-|x_t|^2)|x_\sigma|^2}\over
|x_\sigma|^2}\eqno{(4)}$$ Note that $\Lambda_\pm$
are real by (3).

We would like to have (fairly) sharp conditions
for the existence of time-like curves connecting
points in $Q\cap C$ to $\partial C$
(accessibility). For definiteness we will only
consider accessibility {\it forward} in time
here. The analogs of our results for
accessibility backward in time will be obvious.
Our first result is: \vskip.1in \noindent {\bf
Proposition 2.1.} If any point on $\partial Q$ is
accessible from $\partial C$, then any point in
$Q\cap C$ is accessible from $\partial C$.
\vskip.1in \noindent Proof. Suppose that
$(x_0,t_0)$ is in the interior of $Q$. Since
$\Omega_{t_0}=Q\cap\{t=t_0\}$ is a connected set
with smooth boundary, we can choose a simple path
$x(\sigma)$ parameterized by arc length with
$x(0)=x_0$ and $x(l)\in
\partial \Omega_{t_0}$ such that $x(\sigma)$ is in the interior of
$\Omega_{t_0}$ for $\sigma<l$. Let
$$\sigma_0=\inf\{\sigma\in [0,l]: \cup_{t\in \Bbb R}(x(\sigma),t)\cap
\partial Q\neq\emptyset\}$$
If $\sigma_0=0$, we can just follow the line
$(x_0,t)$ until it hits $\partial Q$. If
$\sigma_0>0$ we proceed as follows. Since the
motion of the boundary is periodic, the vertical
line $(x(\sigma_0),t)$ intersects $\partial Q$ at
a sequence of points $(x(\sigma_0),t_1+nT),\ n\in
\Bbb Z$. Since $\{(x(\sigma),t):\ \sigma\in
[0,\sigma_0),\ t\in \Bbb R\}$ is in the interior
of $Q$, the path $(x(\sigma),t_0-2\sigma),\ 0\leq
\sigma\leq \sigma_0, $ will be time-like and
connect $(x(\sigma_0),t_0-2\sigma_0)$ to
$(x_0,t_0)$ in $Q$. Choose $n$ so that
$t_n=t_1-nT\leq t_0-2\sigma_0$. Since
$(x(\sigma_0),t_n)$ can be connected to $\partial
C$ by a time-like path by hypothesis and
$(x(\sigma_0),t),\ t_n\leq t\leq t_0-2\sigma_0$
is time-like (note that $|x^\prime(\sigma)|=1$),
we see that $(x_0,t_0)$ is accessible from
$\partial C$. \vskip.1in In view of Proposition
2.1 we would like to find conditions implying
that all points on $\partial Q$ are accessible
from $\partial C$. There are always some
accessible points on $\partial Q$: if $x_0$ is a
point on $\partial \Omega_{t_0}$ such that $|x|$
is maximal, then the lines $(x_0\pm
(t-t_0)x_0/|x_0|,t-t_0)$ lie in $Q$ for $t>t_0$
and $t<t_0$ respectively, since the $\partial Q$
is time-like. Hence we have  a sequence of points
$(x_0,t_n)=(x_0,t_0+nT),\ n\in \Bbb Z$, on
$\partial Q$ which are accessible. So we will
look for conditions implying that arbitrary
points on $\partial Q$ can be reached from these
points by time-like curves {\it lying in
$\partial Q$}. Our next result is \vskip.1in
\noindent {\bf Proposition 2.2.} If $\min
\Lambda_+>\max \Lambda_-$, then all points in
$\partial Q$ are accessible in $\partial Q$ from
the points $(x_0,t_n)$.\vskip.1in \noindent
Proof. For $\alpha\in [0,1]$ let
$\sigma_\alpha(t)$ be the solution to
$$\sigma^\prime =\alpha
\Lambda_-(\sigma,t)+(1-\alpha)\Lambda_+(\sigma,t),\
\sigma(t_0)=\sigma_0,$$ where
$x(\sigma_0,t_0)=x_0$. Then
$(x(\sigma_\alpha(t),t),t)$ is time-like for
$\alpha\in [0,1]$ and $\sigma_\alpha(t)$ depends
continuously on $\alpha$. Assuming that
$\sigma_\pm(t)$ have the initial data
$\sigma_\pm(t_0)=\sigma_0$, we have
$\sigma_\alpha(t)=\sigma_-(t)$ when $\alpha =1$
and $\sigma_\alpha(t)=\sigma_+(t)$ when $\alpha
=0$. Hence, it follows by the intermediate value
theorem that $\sigma_\alpha(t)$ takes all values
between $\sigma_-(t)$ and $\sigma_+(t)$ as
$\alpha$ goes from 0 to 1. Thus all of the points
$(x(\sigma, t),t),\
\sigma_-(t)<\sigma<\sigma_+(t),\ t>t_0$ are
accessible from $(x_0,t_0)$.

By the mean value theorem for $t\geq t_0$
$$\sigma_+(t)-\sigma_-(t)
=(t-t_0)(\Lambda_+(\sigma_+(t^*),t^*)-\Lambda_-(\sigma_-(t^*),t^*))$$
for some $t^*$ between $t_0$ and $t$. Thus by the
hypothesis there is a $\delta>0$ such that
$\sigma_+(t)-\sigma_-(t)\geq \delta (t-t_0)$.
Hence, in view of the periodicity of
$x(\sigma,t)$ in $\sigma$, there is a $t_1$ such
that all points on $\{(x,t) \in\partial Q:t\geq
t_1\}$ are accessible from $(x_0,t_0)$. Using the
periodicity of $x(\sigma,t)$ in $t$, it follows
that any point on $\partial Q$ can be reached
from one of the points $(x_0,t_0+nT)$. \vskip.1in

One could conjecture that if the curves
$\sigma_+(t)$ and $\sigma_-(t)$ starting from
$(\sigma_0,t_0)$ are both unbounded, then all
points on $\partial Q$ with $t$ sufficiently
large will be accessible. Unfortunately, this is
not always the case: it is easy to construct
examples (with $\min \Lambda_+=\max \Lambda_-$)
where these curves follow each other so closely
that only a small subset of $\partial Q$ is
accessible from $(x(\sigma_0,t_0),t_0)$. We have
not found a truly sharp hypothesis for
accessibility.

 \vskip.1in The requirement that the
motion be periodic forces the curves
$\sigma_\pm(t)$ to either be unbounded or
asymptotic to periodic orbits as $t\to\pm
\infty$. The precise statement is the following
proposition. \vskip.1in \noindent {\bf
Proposition 2.3.} Assume for simplicity that
$T=1$. Let $\sigma(t)$ be a solution to either
$\sigma^\prime =\Lambda_+(\sigma,t)$ or
$\sigma^\prime =\Lambda_-(\sigma,t)$ for $t \in
\Bbb R$. If $\sigma(1)>\sigma(0)$ and $\sigma(t)$
is bounded above as $t\to \infty$, then
$\sigma(t)$ is asymptotic from below to a
periodic orbit. If $\sigma(1)<\sigma(0)$ and
$\sigma(t)$ is bounded below as $t\to\infty$,
then $\sigma(t)$ is asymptotic from above to a
periodic orbit. If $\sigma(1)=\sigma(0)$, then
$\sigma(t)$ is periodic. \vskip.1in\noindent
Proof. Consider the case when
$\sigma(1)<\sigma(0)$ and $\sigma(t)\geq
\sigma_0>-\infty$ for $t\geq 0$. Since
$\sigma(t+1)$ is also a solution of the equation,
and the solution passing through a point
$(t_0,\sigma_0)$ is unique, it follows that
$\sigma(t)>\sigma(t+1)$ for all $t$. Repeating
this argument one sees that
$\sigma(t)>\sigma(t+1)>\sigma(t+2)>\cdots$ for
all $t$. Defining $w_n(t)=\sigma(t+n)$ we have
decreasing sequence of solutions bounded below by
$\sigma_0$. So
$\lim_{n\to\infty}w_n(t)=w_\infty(t)$ exists for
all $t\geq 0$. Since, letting $F$ denote
$\Lambda_+$ or $\Lambda_-$, we have
$$w_n(t)=w_n(0)+ \int_0^tF(s,w_n(s))ds,$$
the Arzela-Ascoli theorem implies that the
convergence of a subsequence to $w_\infty(t)$ is
uniform on bounded intervals, and hence that
$w_\infty(t)$ is also a solution of
$\sigma^\prime=F(\sigma,t)$. Since
$$w_\infty(0)=\lim_{n\to\infty}\sigma(n)=\lim_{n\to\infty}\sigma(n+1)=w_\infty(1),$$
$w_\infty(t)$ is a periodic solution. Dini's
theorem implies that the convergence of $w_n(t)$
to $w_\infty(t)$ is uniform on $[0,1]$, and for
$t\in [0,1]$ we have
$$|w_\infty(t+n)-\sigma(t+n)|=|w_\infty(t)-w_n(t)|<\epsilon$$
when $n\geq N(\epsilon)$. Hence $\sigma(t)$ is
asymptotic to $w_\infty(t)$ as $t\to \infty$. The
proof for the case $\sigma(1)>\sigma(0)$ is the
same, using increasing sequences in place of
decreasing sequences.

A consequence of Proposition 2.3 is that, when
$\sigma_+(t)$ with $\sigma_+(t_0)=\sigma_0$ is
bounded above and $\sigma_-(t)$ with
$\sigma_-(t_0)=\sigma_0$ is bounded below, there
will be two periodic orbits which may make part
of $\partial Q$ inaccessible (forward and
backward) from $(\sigma_0,t_0)$ along time-like
curves in $\partial Q$. That is what happens in
Stefanov's example. \vskip.2in \centerline {\bf\S
3. Stefanov's Example Revisited} \vskip.1in
Consider a domain with part of its boundary given
by
$$(x_1(\sigma,t),x_2(\sigma,t))= (\sigma,
\phi(\sigma)f(k(2\sigma-t))),\ |\sigma|<M+L,\
k\in \Bbb N$$ Here $|f|\leq 1$ and $f$ is
function of period one, $\phi\in
C^\infty_c(|\sigma|<M+L)$ and $\phi(\sigma)=1$
for $|\sigma|\leq M$. In this construction it
would suffice to have $M=L=2$, but we have kept
the notation $M$ and $L$ to distinguish the
supports of $\phi$  and $\phi^\prime$.

To compute the normal component of $x_t$ we note
$$x_t=(0,k\phi f^\prime)\hbox{ and }x_\sigma=(1,\phi^\prime
f+2k\phi f^\prime).$$ Hence, $|\nu_x\cdot x_t|=
|a|(1+(b+2a)^2)^{-1/2}$, where $a=k\phi f^\prime$
and $b=\phi^\prime f$. From this one can show
that $|x_t\cdot \nu_x|\leq 1/\sqrt 2$, when
$|b|\leq 1$. Since $|f|\leq 1$, it suffices to
have $|\phi^\prime|\leq 1$, and one can arrange
that when $L>1$. Thus with these choices this
portion of $\partial Q$ is time-like.

For this boundary the equation for $\sigma_-(t)$
when $|\sigma_-(t)|\leq M$ is
$$\sigma^\prime_-=\Lambda_-(\sigma_-,t)={2k^2(f^\prime)^2-\sqrt{1+2k^2(f^\prime)^2}\over
1+4k^2(f^\prime)^2}\geq -1.$$ From that one can
see that it is going to be very difficult for
$\sigma_-(t)$ to move to the left. Assume that
$\sigma_-(t_0)=0$.

To check that $\sigma_-(t)$ cannot move very far
to the left, define $w(t)=2\sigma_-(t)-t$. Then
$w(t)$ satisfies the autonomous equation
$$w^\prime=
{-1-2\sqrt{1+2k^2(f^\prime(kw))^2}\over
1+4k^2(f^\prime(kw))^2}=_{\hbox{def}}-F(kw,k).$$
and
$$t-t_0=\int_{w(t)}^{w(t_0)}{dw\over F(kw,k)}=
{1\over k}\int_{kw(t)}^{kw(t_0)}{dz\over
F(z,k)}.$$ For a function of period $1$, writing
$b-a=m+r$ with $m\in \Bbb N$ and $0\leq r<1$, one
has
$$\int_a^bfdz=(b-a)\int_0^1fdz -
r\int_0^1fdz+\int_a^{a+r}fdz.$$ Applying this
with $f=(kF(z,k))^{-1}$
 gives
$$t-t_0=(w(t_0)-w(t))H_0 -{r\over k}H_0 +{1\over k}\int_{kw(t)}^{kw(t)+r}{dz\over F(z,k)},$$
where $0 \leq r<1$ and $H_0=\int_0^1dz/F(z,k)$.
Substituting $2\sigma_-(t)-t$ for $w(t)$ and
solving gives
$$2\sigma_-(t)=(1-{1\over H_0})(t-t_0)-{r\over k}+{1\over
kH_0}\int_{kw(t)}^{kw(t)+r}{dz\over F(z,k)}\geq
(1-{1\over H_0})(t-t_0)-{1\over k},$$ since
$F(z,k)$ does not change sign. Thus, assuming
$M>1/2$, $\sigma_-(t)$ will never reach $-M$, if
$H_0>1$. We have

$$H_0=$$
$$\int_0^1{1+4k^2(f^\prime(z))^2\over 1+2\sqrt{1+2k^2(f^\prime(z))^2}}dz>\int_0^1{1\over
2}\sqrt{1+2k^2(f^\prime(z))^2}dz>{k\over \sqrt
2}\int_0^1|f^\prime(z)|dz.
 $$
So  $|H_0|>1$ when $k\int_0^1|f^\prime(z)|dz
>\sqrt 2$. This determines our choice of $k$ and
shows that for this example points on $\partial
Q$ with $\sigma =-M$ are inaccessible (forward)
in the boundary from points with $\sigma =M$.

To turn this into an example of a domain
$Q\subset \Bbb R^2_x\times \Bbb R_t$ with points
on $\partial Q$ inaccessible from $\partial C$,
we need to specify the rest of $\partial Q$ in a
way that forces any curve in $Q$ reaching
$x=(M+L,0)$ to follow the boundary constructed
above very closely. Here we use the original idea
in [St]: we add another boundary curve below, but
very close to the boundary just constructed, so
that any curve in $Q$ reaching $x=(-M-L,0)$ must
pass through the narrow \lq\lq channel" between
these curves.

 Let $\nu(\sigma,t)$ be the unit normal (directed downward) to the curve $x(\sigma,t)$ and
 consider
 $$x(\sigma,t,\eta)=x(\sigma,t)+\eta\nu(\sigma,t)$$
Since the curvature of the curve traced by
$x(\cdot, t)$ is bounded for all $t$, there is a
$ \delta>0$ such that  $(\sigma,\eta)$ are
coordinates on $\{x: |x_1|<M+L,\
x_2(x_1,t)-\epsilon <x_2\leq x_2(x_1,t)\}$ for
$\epsilon$ sufficiently small. We are going to
take $x_2=x_2(x_1,t)-\epsilon$ as the boundary of
the lower side of the channel, but we will be
taking $\epsilon$ smaller later in the argument.

A curve of speed less than one in the channel
$x(\sigma(t),t,\eta(t))$ satisfies
$$1\geq |\dot \sigma x_\sigma +x_t+\dot\eta \nu +\eta \dot \nu|^2$$ $$=|x_\sigma|^2\dot \sigma^2+ |x_t|^2+
\dot  \eta^2 +\eta^2|\dot \nu|^2+2\dot \sigma
x_\sigma\cdot x_t +2\eta\dot \sigma x_\sigma\cdot
\dot \nu +2\dot \eta \nu\cdot x_t +2\eta x_t\cdot
\dot \nu,\eqno{(5)}$$ where we used $\nu\cdot
x_\sigma=\nu\cdot \dot \nu=0$ since $\nu$ is the
unit normal. Since $|\nu\cdot x_t|<1$, we have
$\dot \eta^2+2\dot \eta \nu\cdot x_t+1>0$ and can
rewrite (5) as
$$2\geq |x_\sigma|^2\dot \sigma^2+ |x_t|^2+
 +2\dot \sigma x_\sigma\cdot x_t -O(\eta),\eqno{(6)}$$
 where the $O(\eta)$ term does not depend on $\dot \eta$. Solving (6) for $\dot \sigma$ gives
 $$\tilde \Lambda_-(\sigma, t)\leq \dot \sigma \leq \tilde \Lambda_+(\sigma,t),\ \tilde
 \Lambda_\pm ={-x_\sigma\cdot x_t\pm \sqrt{(x_\sigma\cdot
x_t)^2+(2+O(\eta)-|x_t|^2)|x_\sigma|^2}\over
|x_\sigma|^2}.$$ Now suppose that the $O(\eta)$
term is less than 1. Then the argument given
earlier shows that $\sigma (t)$ cannot reach $-M$
for any $t>0$ when $k$ is sufficiently large (in
this case it suffices to have
$k\int_0^1|f^\prime(z)|dz >\sqrt 6$). Having
chosen that $k$ we then take $\delta$
sufficiently small that the $O(\eta)$ term is
less than 1, and finally choose $\epsilon$ so
that the channel is entirely in the region where
$0\leq \eta <\delta$. Thus no time-like curve can
pass through the channel and the points in $Q$
with $x_1=-M-L, -\epsilon <x_2<0$ are
inaccessible (forward) for all time.
 \vskip.1in\noindent {\bf Remark}. In [St] Stefanov
assumed that the rest of $\partial Q$ was defined
so that the intersection of the cylinder
$\{|x-(-M-L,0)|<2M+2L\}\times \Bbb R_t$ with $Q$
only contains the channel. Then one can use the
domain of dependence theorem of Inoue [In] to
show that, if a solution to the forward problem
were nonzero near $x=(-M-L,0)$ at some time, then
there would necessarily be a time-like path
through the channel. Hence, by contradiction,
$((-M-L,0),t)$ is outside the domain of
dependence of $\partial C$ for all $t$.\vskip.2in

  \centerline{\bf \S 4.
Determination of $Q$ from Cauchy Data}

\centerline{\bf for General Hyperbolic
Equations}\vskip.1in
 \noindent {\bf \S 4.1.}  In this section we consider the more general hyperbolic equation $Lu=0$ from the Introduction with $$Lu=
\partial^2_t u-2a(t,x)\cdot \nabla_x
 \partial_tu -\nabla_x\cdot A(t,x)\nabla_x u
-L_1(t,x,\partial_t,\partial_x)u,$$  where $L_1$
is a differential operator of order one. All
coefficients are assumed to be real, bounded and
smooth on ${\Bbb  R}_t\times {\Bbb R}^n_x$, and
real analytic in $t$.  $L$ is strictly hyperbolic
with respect to $t$ if for $\xi \neq 0$
$$0=p_2(x,t,\xi,\tau)=_{\hbox{def.}}\tau^2-2\tau
a\cdot\xi-\xi\cdot A\xi$$ has distinct real roots
$$\tau_\pm(x,t,\xi)=a\cdot \xi\pm \sqrt{(a\cdot  \xi)^2+\xi\cdot
A\xi}.$$ We make the stronger hypothesis that
$A(x,t)\geq \delta I,\ \delta>0$, for all
$(t,x)\in \Bbb R^n_x\times \Bbb R_t$. Hence
$\tau_+$ and $\tau_-$ have opposite signs. This
has the following interpretation in
pseudo-riemannian geometry. Writing
$p_2(x,t,\xi,\tau)$ as a quadratic form
$(\xi,\tau)\cdot B(x,t)(\xi,\tau)$, the dual form
on tangent vectors is $(v_x,v_t)\cdot
B^{-1}(x,t)(v_x,v_t)$. One says that a curve in
space-time is \lq\lq time-like" if its tangent
vector $v=(v_ x ,v_tt)$ satisfies $v\cdot
B^{-1}(x,t)v>0$. The hypothesis $A>0$ is
equivalent to assuming that  the time curves
$\gamma(t)=(x_0,t)$ are time-like for $L$ (see \S
4.2).

We consider solutions of $Lu=0$ in a
time-dependent (open) domain $ Q$ in ${\Bbb
R}_x^n\times {\Bbb R}_t$ where the  boundary
$\partial Q$ is smooth and uniformly time-like
for $L$, i.e. there is a $\delta>0$ such that
$p_2(x,t,\xi,\tau)\leq -\delta |(\xi,\tau)|^2$
when $(\xi,\tau)$ is normal to $\partial Q$ at
$(x,t)$. We also assume that for each $t\in \Bbb
R$ the set $\Omega_t=\{x: (x,t)\in Q\}$ is a
connected exterior domain in $\Bbb R^n$, and the
complement of $Q$ is contained in the cylinder
$C=\{(t,x): |x|<\rho\}$. For positive results we
will need the following additional hypothesis on
$Q$. We assume that we have diffeomorphisms
$\Psi^t(y)$ with the following properties:
\vskip.1in (i) for each $t$, $\Psi^t(y)$ maps
$\Bbb R^n$ onto $\Bbb R^n$ taking $\Omega_0$ onto
$\Omega_t$ (and $\partial\Omega_0$ onto $\partial
\Omega_t$), \vskip.1in (ii)$\Psi^0(y)=y,\ y\in
\Omega_0$, and  $\Psi^t(y)=y$ near $\partial C$
for all $t\in \Bbb R_t$, \vskip.1in (iii)
$(\Psi^t(y),t)$ is time-like for $L$ for all
$y\in \Omega_0$. We assume that this holds
uniformly in the sense that
$(\partial_t\Psi^t,1)\cdot
B^{-1}(\Psi^t,t)(\partial_t\Psi^t,1)\leq
-\delta<0$ for $(y,t)\in \Omega_0\times \Bbb R_t$
and \vskip.1in (iv) the Jacobian matrix of
$\Psi^t$ satisfies $ ||\partial_y\Psi^t(y)||\leq
M<\infty$ for all $(y,t)\in \Omega_0\times \Bbb
R_t$. \vskip.1in \noindent In [St] Stefanov gave
a construction for $L=\partial^2_t-\Delta$ of
$\Psi^t$ satisfying (i)-(iii) for {\it any} $Q$
with time-like boundary and complement in $C$ as
the solution of $\dot x=v(x,t),\ x(0,y)=y$ for a
suitable vector field $v(x,t)$ on $Q$, tangent to
$\partial Q$. This can be generalized to the
setting here (see \S 4.2).  So the additional
hypothesis here is really (iv). Now we can state
\vskip.1in \noindent {\bf Theorem 4.1.} Suppose
that $R$ is another connected domain with
time-like boundary and complement contained in
$C$. Let $\Gamma$ be an open subset of
$\{|x|=\rho\}$. If the Cauchy data $\Cal K(Q)$
and $\Cal K(R)$ are identical on $\Gamma\times
\Bbb R_t$, then $R=Q$. \vskip.1in \noindent
Proof. We will work in the domain $\hat
Q=\Omega_0\times \Bbb R_t$, pulling $L$ back to
$\hat L$ on this domain by the mapping
$(x,t)=\Phi(y,t)=(\Psi^t(y),t)$.

Suppose that $R$ is a second domain as in the
statement of the theorem. If $R\neq Q$, then
there will be either  boundary points of $R$ in
the interior of $Q$ or boundary points of $Q$ in
the interior of $R$. We begin with $(x_0,t_0)$
which is a boundary point of $R$ in the interior
of $Q$ and let $(y_0,t_0)$ be its pre-image under
$\Phi$. By the assumption that $\Omega_{t_0}$  is
connected,  we can choose a  smooth,
non-self-intersecting path $y(\sigma),\ 0\leq
\sigma\leq l$ in $\Omega_{t_0}$,
 such that $|y^\prime(\sigma)|=1$, $y(0)=y_0$ and
$y(l)\in \Gamma$. For convenience later we choose
$y(\sigma)$ so that $y^\prime(l)$ is radial and
hence normal to $\partial C$.

 Now we would like to choose a constant $a>0$ so that
$\gamma_+(t)=(y(a(t-t_0)),t)$ is time-like for
$\hat L$ for $t_0\leq t\leq t_0+l/a$. Hence we
want $\dot \gamma_+\cdot \hat
B^{-1}\dot\gamma_+>0$, where $\hat B$ is the
quadratic form associated with $\hat L$. Note
that
$$\hat
B^{-1}(y,t)=(\partial_{y,t}\Phi(y,t))^TB^{-1}(\Phi(y,t))(\partial_{y,t}\Phi(y,t)).$$
Thus hypotheses (iii) and (iv) imply that
$$|(ay^\prime(\sigma)),1)\cdot
\hat B^{-1}(y,t)(ay^\prime(\sigma),1)-(0,1)\cdot
\hat B^{-1}(0,1)|\leq Ca$$ uniformly for
$(y,t)\in \hat Q$ and $\sigma\in [0,l]$. This is
the crucial use of (iv): there is no choice of
$\Phi$ which will make this estimate true in
Stefanov's example. Since hypothesis (iii) also
implies that the vector $(v_x,v_t)= (0,1)$ is
uniformly time-like for $\hat L$, we can choose
the constant $a$ so that $\gamma_+$ is time-like.
Likewise, $\gamma_-(t)=(y(-a(t-t_0)),t)$ is
time-like for $t_0-l/a\leq t\leq t_0$.

Consider the two dimensional surface $S_0$ given
by
$$S_0=\{(y(a(t-t_0)),s),\ |s-t_0|\leq |t-t_0|\leq l/a\}.$$
$S_0$ is roughly triangular with boundary curves
$\gamma_+$, $\gamma_-$ and
$\gamma_0(t)=(y(l),t)),\ |t-t_0|\leq l/a$.
Clearly with our hypotheses  it possible that
$S_0\cap
\partial \hat R\neq \emptyset$, where $\hat R $ is the pull-back of $R$ under $\Psi$.
 However, since $S_0$ is compact and the points on $S_0$ could be parameterized by
$(\sigma,s)$, where $\sigma =a|t-t_0|$, we can
choose $(y_1,t_1)\in S_0\cap
\partial \hat R$ where $\sigma$ assumes its maximum, $\sigma_1$. Then we
repeat the construction of $S_0$, using $t_1$ in
place of $t_0$, $y(a(t-t_1)+\sigma_1)$ in place
of $y(a(t-t_0))$, and $y(-a(t-t_1)+\sigma_1)$ in
place of $y(-a(t-t_0))$. Note that we can take
the new $S_0$ to be a subset of the original
$S_0$. After this correction, we go back to the
original notation, relabeling $(y_1,t_1)$ as
$(y_0,t_0)$. Hence we can now assume that $S_0$
is contained in the interior of $\hat Q$ and
intersects the boundary of $\hat R$ only at
$(y_0,t_0)$. Note also that, since $\partial \hat
R$ is time-like, taking $a$ slightly {\it larger}
we can assume that $\gamma_+$ and $\gamma_-$ are
not tangent to $\partial \hat R$ at $(y_0,t_0)$.
In what follows we will consider the surface
$\Sigma_\epsilon =\partial \hat R\cap
\{(y,t):|(y-y_0,t-t_0)|<\epsilon\}$ with
$\epsilon$ small enough that $\Sigma_\epsilon$
intersects the sphere of radius $\epsilon$ around
$(y_0,t_0)$ transversally.

In this proof we will use recent extensions of
Holmgren's uniqueness theorem to reach a
contradiction to the existence of $(x_0,t_0)$.
This requires an increasing sequence of domains
in $\hat Q\cap \hat R$ with smooth time-like
boundaries which connect $\Gamma\times \Bbb R_t$
to a domain whose boundary contains a portion of
$\Sigma_\epsilon$. In what follows we construct
such a sequence so that the boundary  of the
final domain intersects $\partial \hat R$ only at
$(x_0,t_0)$. However, it will be clear that we
could have performed the same construction with
$(x_0,t_0)$ replaced by any nearby point on
$\Sigma_\epsilon$. This will give us a
sufficiently large set of domains in which to use
the uniqueness theorem, and reach a
contradiction.

 We begin the construction by introducing two
dimensional surfaces $S_\sigma$ contained in
$S_0$
$$S_\sigma =\{(y(a(\sigma)(t-t_0))+\sigma,s),\ |s-t_0|\leq |t-t_0|\leq
l/a\},\ 0<\sigma \leq l,$$ where  $a(\sigma)=
(1-\sigma/l)a$. Note that, since $0\leq
a(\sigma)\leq a$, the upper and lower boundary
curves of $S_\sigma$, $\gamma^\sigma_\pm$, are
time-like. The surfaces $S_\sigma$ share the
boundary curve $\gamma_0$ in $\partial C$, and
have the points $(y(\sigma),t_0),\ 0\leq \sigma
\leq l$, as \lq\lq vertices". Note that the
surfaces $S_\sigma$ are nested: $S_\sigma\subset
S_{\sigma^\prime}$ when $\sigma>\sigma^\prime$.

Next, we consider the bounded regions $D_\pm$
bounded by $\partial C$ and the parametric
surfaces $B_\pm$ defined by
$$\{ (y(s),t_0\pm s/a)+r(s)\omega: 0<s<l,\ \omega\in \Bbb
S^n\cap\pi(s)\},$$ where $\pi(s)$ is the plane
through $(y(s),t_0)$ perpendicular to
$(y^\prime(s),0)$. Hence $B_+$ and $B_-$ are
unions of $(n-1)$-dimensional spheres in the
$n$-dimensional planes perpendicular to the curve
$(y(s),0),\ 0<s\leq l$, of varying radii with
centers on $\gamma_+$ and $\gamma_-$
respectively. In order for these regions to lie
in $\hat Q\cap \hat R$ and have smooth boundary
$r(s)$ must be small. We choose $r(s)$ small,
tending to zero as $s$ goes to zero, and
increasing monotonically with $s$. However, we
keep it small enough that $D_\pm\cap
\partial C\subset \Gamma\times \Bbb R_t$.

Next we take $D_0$ to be the union of the convex
hulls of the intersections of $D_\pm$ with the
planes $\pi(s)$ for $0<s\leq l$. In other words
$D_0$ is the union of a family of \lq\lq stadium
domains", the convex hulls of sets consisting of
two spheres. The part of $\partial D_0$ which is
not in $B_-\cup B_+$ consists of vertical lines
(\lq\lq vertical" means parallel to the $t$-axis)
and hence is time-like. Any point in the boundary
of $D_0$ which is in $B_+\cup B_-$ must have the
form given in the parametrization above with the
$t$ component of $\omega$ nonnegative on $B_+$
and nonpositive on $B_-$. Hence, taking the curve
though such a point varying $s$ in the
parametrization and holding $\omega$ constant,
one gets a curve with a time-like tangent. Hence
the boundary of $D_0\cap C$ is time-like (see \S
4.2 for a proof of the basic result that a
hypersurface containing a time-like curve will be
time-like at all points on that curve).

To construct an exhaustion of $D_0$ by increasing
domains with time-like boundaries we repeat the
preceding replacing, $\gamma_\pm$ by
$\gamma_\pm^\sigma$. We construct  $B^\sigma_\pm$
as the union of spheres in the space-time planes
perpendicular to $(y(s),t_0),\
 \sigma\leq s\leq l$ with the same $r(s)$ as before. However, when
 we take the convex hulls of the regions bounded by these surfaces in the planes $\pi(s)$,
 they end in the disk of radius
 $r(\sigma)$ centered at $(y(\sigma),t_0)$ in
 the  plane perpendicular to $(y^\prime(\sigma),0)$.  However, we know that the tangent planes to the
 boundary of this set remain strictly time-like when one approaches the disk from $s>\sigma$. Hence there is
 no difficulty building $D_\sigma$ by adding a \lq\lq cap" to the region
 ending in the disk so that $D_\sigma$
 time-like boundary and contains $(y(s),t_0)$ for $\sigma>s>\sigma
 -\delta(\sigma)>0$. Moreover, one can add the caps to that
 $D_\sigma$ is monotonically decreasing with $\sigma$.

The boundaries of the domains $D_\sigma$ are
$C^1$, failing to be smooth where the vertical
lines from the convex hulls touch $B^\sigma_\pm$.
However, one can smooth them near these points
preserving the time-like boundaries and
monotonicity of the $D_\sigma$'s.

 To construct the domain $D_0$ and its exhaustion in
the case that $(y_0,t_0)$ is a boundary point of
$\hat Q$ in the interior of $\hat R$, one
proceeds in the same way, defining
$\Sigma_\epsilon =\partial \hat Q\cap
\{|(y,t):|(y-y_0,t-t_0)|<\epsilon\}$. Note that
the only case of $R\neq Q$ where there will be no
interior points of $Q$ which are boundary points
of $R$ is $Q\subset R$. Hence without loss of
generality we can assume that $S_0\cap
\partial \hat R=\emptyset$ here, omitting the consideration of
$(x_1,t_1)\in S_0\cap \partial \hat R$ required
in the preceding case.

This is the construction needed for the use
unique continuation theorems in the case that
$\partial D_0\cap \partial\hat R=\{(x_0,t_0)\}$.
For $(x^\prime,t^\prime)\in \Sigma_\epsilon$
sufficiently close to $(x_0,t_0)$ one can repeat
the construction replacing $x_0$ by $x^\prime$
and $t_0$ by $t^\prime$ at all places where they
appear. There is one slightly subtle point in
this argument: in order to replace $S_0$ by a
small perturbation $S^\prime$ without hitting new
points of $\partial \hat R$ the condition that
$\partial \hat R$ is not tangent to $\gamma_\pm$
at $(x_0,t_0)$ must be used. We denote the domain
$D_0$ ending at $(x^\prime,t^\prime)$ by
$D_0(x^\prime,t^\prime)$. \vskip.1in
 Our assumption ${\Cal
K}(R)={\Cal K}(Q)$ on $\Gamma\times \Bbb R_t$
implies that solutions to the forward problem in
$Q\cap C$
$$Lu=0 \hbox{ in } Q\cap C,\ u=0 \hbox{ on }
\partial Q,\ u=f \hbox{ on }
\partial C,\hbox{ and }u=0 \hbox{ when }t<<0,$$
and the same problem with $Q$ replaced by $R$,
have the same Cauchy data on $\Gamma\times \Bbb
R_t$. This will make them identical on the sets
$\Phi(D_0(x^\prime,t^\prime))$. To prove that
note first that, since the boundaries of the
$D_\sigma$'s are time-like for $\hat L$, the
boundaries of their images under $\Phi$ are
time-like for $L$. Thus we have an exhaustion of
$\Phi(D_0(x^\prime,t^\prime))$ regions with
time-like boundaries which intersect $\partial C$
in fixed subset of $\Gamma\times \Bbb R_t$. Thus,
 assuming that $u^f_Q-u^f_R$ does not vanish
identically in $\Phi(D_0(x^\prime,t^\prime))$,
there is a last $\sigma$ such that it vanishes on
$D_\sigma(x^\prime,t^\prime)$. Since we have
assumed that the coefficients of $L$ are analytic
in $t$, the unique continuation theorems of
Robbiano-Zuily and Tataru (see [RZ] and [T]) give
a contradiction, and we conclude that
$u^f_Q-u^f_R$ vanishes on
$\Phi(D_0(x^\prime,t^\prime))$. Note that for
$L=\partial_t^2-\Delta$ this step only requires
Holmgren's theorem. Since we have this conclusion
for all $(x^\prime,t^\prime)$ in a neighborhood
$\Sigma$ of $(x_0,t_0)$ in $\Sigma_\epsilon$
(meaning $(x_0,t_0)\in\Sigma\subset
\Sigma_\epsilon$), we conclude in conclusion that
$u^f_Q-u^f_R$ vanishes on
$$D=\cup_{(x^\prime,t^\prime)\in \Sigma}D_0(x^\prime,t^\prime).$$

 Let $G_Q(x,z,t,s)$ and $G_R(x,z,t,s)$ be the
backward fundamental solutions (see [I]) for
$L^*$, the adjoint of $L$, in $Q\cap C$ and
$R\cap C$ respectively, i.e.
$$L^*G_Q=L^*G_R=\delta(x-z,t-s),\
G_Q=G_R\equiv 0 \hbox{ when } t>s,\hbox{ and }$$
$G_Q=0$ on $\partial Q\cup \partial C$, $G_R=0$
on $\partial R\cup \partial C$. Given $g\in
C_c^\infty(\Phi(D))$, let $v^g_Q$ and $v^g_R$ be
$G_Q$ and $G_R$ applied to $g$ respectively.
Hence $L^*v_Q^g=L^*v_R^g=g$, $v_Q^g$ and $v_R^g$
vanish on $\partial Q\cup \partial C$ and
$\partial R\cup
\partial C$ respectively, and $v_Q^g=v_R^g=0$ for $t$ sufficiently
large. We have
$$\int_{Q\cap C} g(x,t)u^f_Q(x,t)dxdt
=\int_{\partial C}f(x,t)\nu\cdot
A(x,t)\nabla_xv^g_Q(x,t)dtdS, \hbox{ and
}$$$$\int_{R\cap C} g(x,t)u^f_R(x,t)dxdt
=\int_{\partial C}f(x,t)\nu\cdot
A(x,t)\nabla_xv^g_R(x,t)dtdS,$$ where $\nu=x/|x|$
is the normal to $\partial C$.

 Since we know that $u^f_Q=u^f_R$ on $\Phi(D)$ for all
$f$ , it follows that
$$\nu\cdot A(x,t)\nabla_x v^g_Q (x,t)=\nu\cdot A(x,t)\nabla_x v^g_R(x,t)$$ when $(x,t)\in
\partial C$. However, by construction $v^g_Q=v^g_R=0$ on $\partial C$
and $L^*(v^g_Q-v^g_R)=0$ in $Q\cap R$. Thus we
can use the unique continuation argument which
showed $u^f_Q=u^f_R$ in $\Phi(D)$ again -- this
time for $L^*$ instead of $L$ -- to conclude that
$v^g_Q=v^g_R$ on $\Phi(D)$. Finally, since $g$
was an arbitrary function in
$C_c^\infty(\Phi(D))$, we have$$G_Q(
x,t,z,s)=G_R(x,t,z,s) \hbox{ for all
}(x,t),(z,s)\in \Phi(D)\eqno{(7)}.$$ Taking
$(z,s)$ in $\Phi(D)$ sufficiently close to
$(x_0,t_0)$ in (7) leads to a contradiction to
the propagation of singularities for these
fundamental solutions. The wave front set of
$G_Q(x,t,z,s)$, considered as a distribution in
$(x,t)$, consists of all backward null
bicharacteristics for $L$ passing over $(z,s)$.
Since $G_R$ is constructed with the boundary
condition $u=0$ on $\partial R$, the
corresponding singularities for it are reflected
by $\Phi(\Sigma)$ (see [H\"o]). We will take
$(z,s)$ in $D$ close enough to $\Phi(\Sigma)$
that some singularities will reflect. Then we
have a contradiction to the equality of the
fundamental solutions. This contradiction
completes the proof, and we conclude that $Q=R$.
\vskip.1in \noindent {\bf \S 4.2.} In this
section we give three results from
pseudo-riemannian geometry mentioned in \S 4.1.

First we show that $(x_0,t)$ will be a time-like
curve for $L$ if and only if $A$ is positive
definite. Let $B$ be the matrix of the quadratic
form $p_2(x,t,\xi,\tau)$ as before. We have
$(0,1)\cdot B^{-1}(0,1)=(B^{-1})_{n+1,n+1}$.
Since
$$ B=\left(\matrix -A&-a\\-a&1\endmatrix \right),$$ the formula for the inverse
gives
$$(B^{-1})_{n+1,n+1}={\hbox{det}(-A)\over \hbox{det}(B)}=(-1)^n{\hbox{det}(A)\over \hbox{det}(B)}$$
Since $B$ has one positive and $n$ negative
eigenvalues $(-1)^n\hbox{det}(B)$ is positive.
Since the quadratic form $w\cdot Aw+(w\cdot a)^2$
is positive definite, $A$ has at most one
nonpositive eigenvalue. Thus $(B^{-1})_{n+1,n+1}$
is positive if and only if $A$ is positive
definite. \vskip.1in Next we show that, if
$v\cdot B^{-1}v>0$ and $w$ is a nonzero vector
such that  $v\cdot w=0$, then $w\cdot Bw<0$. Note
that this implies that a smooth surface of
codimension one containing a time-like curve will
be time-like at all points on the curve. To see
that $w\cdot Bw<0$ let $Z_0$ be the normalized
eigenvector of $B$ belonging to the positive
eigenvalue $\lambda_0$. Then we have the
orthogonal decomposition of $\Bbb R^{n+1}$ into
invariant subspaces for $B$, $\Bbb R^{n+1}=<Z_0>
\oplus <Z_0>^\perp$. Let $-C$ denote the
restriction of $B$ to $<Z_0>^\perp$. Since $B$ is
negative definite on $<Z_0>^\perp$, $C$ is
positive definite. We decompose $v$ and $w$ with
respect to this orthogonal decomposition as
$v=v_0Z_0+\tilde v$ and $w=w_0Z_0+\tilde w$. If
$w_0=0$, we have $w\cdot Bw<0$. Since $v\cdot
w=0$, $\tilde v=0$ implies $w_0=0$, and we have
$w\cdot Bw<0$ again. In the other cases we have
$$ v_0^2w_0^2=|\tilde v\cdot \tilde w|^2=
|C^{-1/2}\tilde v\cdot C^{1/2}\tilde w|^2\leq
(\tilde v\cdot C^{-1}\tilde v)(\tilde w\cdot
C\tilde w)\eqno{(8)}$$ by the Cauchy-Schwarz
inequality. Since $v\cdot B^{-1}v>0$, we have
 $v_0^2>\lambda_0\tilde v
\cdot C^{-1}\tilde v$. Substituting this for
$v_0^2$ in (8), we have $\lambda_0w_0^2<\tilde
w\cdot C\tilde w$ which is equivalent to $w\cdot
Bw<0$.
 \vskip.1in
 Next  we show that one can construct
$\Phi$ satisfying (i)-(iii) for any $Q$ with
complement contained in $C$ and time-like
boundary. We will use an analog of the flow
$F_{t,s}(x)$ constructed in \S 3 of [St], adapted
to the general hyperbolic operator $L$. To make
all estimates uniform as $|t|\to\infty$ we assume
that the coefficients of $L$ and the unit normals
to $\partial Q$ have uniformly bounded
derivatives, that $A(x,t)$ is uniformly positive
definite, and that $\partial Q$ is uniformly
time-like. To construct $\Phi(y,t)$ we are going
to define a vector field $v(x,t)$ on $ Q$ and
solve
$${dF\over dt}=v(F,t),\ F(0,y)=y$$
to obtain a diffeomorphism of $\Omega_0$ onto
$\{F(t,y): y\in\Omega_0\}$. If the  vector field
$(v(x,t),1)$ is tangent
 to $\partial Q$, then  $\{F(t,y):
y\in\Omega_0\}$  will be $\Omega_t$ (see [H,
Chpt. 8] for details). Hence, the mapping
$$\Phi(y,t)=(F(t,y),t)$$  is a diffeomorphism of the cylinder $\Omega_0\times\Bbb
R_t$ onto $Q$.  So to satisfy conditions
(i)-(iii) we just need to show that we can choose
$v(x,t)$ vanishing near $\partial C$ so that
$(v(x,t),1)$ is  time-like for $L$. Writing
$p_2(x,t,\xi,\tau)=(\xi,\tau)\cdot
B(x,t)(\xi,\tau)$ as before, we know that $B$ has
one positive and $n$ negative eigenvalues. Hence,
letting $\hat e_0$ be a unit eigenvector for the
positive eigenvalue $\lambda_0$, $B$ is negative
definite on the orthogonal complement of $\hat
e_0$. Since we are assuming that $\partial Q$ is
time-like, its normal $\nu$ satisfies $\nu\cdot
B\nu<0$. We write $\nu=a_0\hat e_0 + w$ with
$w\cdot \hat e_0=0$, and normalize $\nu$ by
requiring $w\cdot Bw=-1$. So with this
normalization $\nu\cdot B\nu<0$ is equivalent to
$a_0^2\lambda_0<1$. We will look for a time-like
vector $d$ tangent to $\partial Q$ in the from
$$d=\hat e_0+z,\ \hat e_0\cdot z=0.$$
The choice $z=a_0Bw$, where $w$ is the vector in
the representation for $\nu$ above, makes $d\cdot
\nu=0$, and turns out to make $d$ time-like as
well: we have
$$d\cdot B^{-1}d=\lambda_0^{-1}+a_0^2Bw\cdot
B^{-1}Bw=\lambda_0^{-1}-a_0^2>0.$$ We make this
choice of $d(x,t)$ on $\partial Q$. Since the
time component of $d$ is uniformly bounded away
from zero, we can normalize it to $(v(x,t),1)$ as
required.

Since we assume that the derivatives of the
coefficients of $L$ and the derivatives of the
unit normals to $\partial Q$ are globally
bounded, there is a neighborhood
$$\Cal N_\delta =\{(x,t)+s\nu(x,t)), (x,t)\in \partial  Q,\
0\leq  s\leq \delta\}$$ such that the extension
of $v(x,t)$ given by
$$v((x,t)+s\nu(x,t))=v(x,t)$$
remains in the time-like cone for $L$. Hence we
can smoothly deform $v$ inside the time-like cone
to (0,1) inside $N_\delta$. This assures that
$\Phi$ satisfies (i)-(iii).

\vskip.1in \noindent {\bf \S 4.3.} In this
section we discuss four examples where moving
boundaries are determined uniquely by Cauchy
data. \vskip.1in \noindent {\bf 1. Rigidly moving
bodies.} Take $L=\partial_t^2-\Delta$ and $\Gamma
=\partial C$. We assume that $Q$ is generated by
the rotation and translation of a rigid body
$\Cal B\subset \{|y|<\rho_0<\rho\}$. Hence
$$\partial Q=\{(O(t)y+l(t),t): y\in\partial \Cal
B\}.$$  $O(t)$ is the rotation (a real orthogonal
matrix of determinant 1), and $l(t)$ is the
translation. Without loss of generality we assume
$O(0)=I$ and $l(0)=0$. To keep $\partial Q$ in
$\{(x,t): |x|<\rho\}$ we assume that $|l(t)|<
\rho-\rho_0-\epsilon$ for all $t$. Finally we
assume that the derivatives $O^\prime(t)$ and
$l^\prime(t)$ are small enough that we have
$$|O^\prime(t)y|+|l^\prime(t)|\leq c <1$$
when $|y|\leq \rho$. The last assumption makes
$\partial Q$ uniformly time-like. Note that, when
$O(t)\equiv I$, this assumption reduces to
$|l^\prime(t)|\leq c<1$ which is necessary as
well as sufficient for $\partial Q$ to be
uniformly time-like.

In this setting we replace hypothesis (ii) on
$\Psi^t(y)$ by \vskip.1in \noindent
(ii')$\Psi^0(y)=y,\ y\in \Omega_0$, and
$\Psi^t(y)$ maps $\partial C$ onto $\partial C$
for all $t\in \Bbb R_t$. \vskip.1in \noindent
Since $O(t)$ is allowed to continue twisting in
the same direction for all time -- think of a
rotation in two space dimensions -- it appears
difficult to construct $\Psi^t$ satisfying
conditions (ii) and (iv) simultaneously. However,
since we are taking $\Gamma$ to be the whole
boundary of $C$, $\Psi^t$ satisfying (i), (ii'),
(iii) and (iv) will suffice for the proof of
Theorem 4.1. Note that, even though $\Psi^t(y)$
is not the identity on $\partial C$, equal Cauchy
data in the $x$-coordinates correspond to equal
Cauchy data in the $y$-coordinates. The
construction of this $\Psi^t$ can be done as
follows.

Note first that $y\to Oy+\beta(|y|)l$ is a
diffeomorphism on $|y|>0$ when $\beta(s)$ is a
smooth real-valued function satisfying
$|l||\beta^\prime(s)|<1$. With this in mind
choose a smooth $\beta(s)$ satisfying
$\beta(s)=1$ for $s\leq \rho_0$ and $\beta(s)=0$
for $s\geq \rho$ such that
$|l(t)||\beta^\prime(s)|<1$ for all $t$. This
will be possible for some $\epsilon>0$, since we
have  $|l(t)|<\rho-\rho_0-\epsilon$ for all $t$,
as assumed above. One checks easily that with
this choice of $\beta$
$$\Psi^t(y)=O(t)y+\beta(|y|)\vec l(t)$$
satisfies all the requirements.

 \noindent {\bf  2. Even periodic
motion.} Suppose that $Q$ is invariant under both
the maps $t\to t+1$ and $t\to -t$. So for each
$n\in \Bbb Z$ and $t\in [0,1/2]$ we have
$\Omega(n+t)=\Omega(n+1-t)$. This makes it
possible to define $\Psi$ as follows: for $t\in
[0,1/2]$ define $\Psi(y,t)=(F(t,y),t)$ where
$F(t,y)$ is the mapping from \S 4.2. For $1/2\leq
t\leq 1$ use $\Psi(y,t)=F(1-t,y)$, and then
continue periodically. The resulting mapping
satisfies (i)-(iv), but has jumps its time
derivative. However, this does not affect the
argument. Note that both of the one sided
derivatives $\Phi_t^+$ and $\Phi_t^-$ are
time-like. It is interesting to compare this with
the example in \S 3: the back and forth motion
here rules out inaccessible regions. \vskip.1in
\noindent {\bf 3. \lq\lq Slow and uniform"
periodic motion}. Consider domains $Q\subset \Bbb
R^n_x\times \Bbb R_t$ with $\partial Q$ given by
$x(y,t),\ y\in
\partial \Omega_0,$ where $x(y,t+1)=x(y,t)$. We take $L=\partial_t^2-\Delta$ and make the
following assumptions for all $(y,t)\in
\partial\Omega_0\times \Bbb R_t$
$$ |x_t(y,t)| \leq \epsilon_0 \hbox{ and }| D_u x_t(y,t)|\leq \epsilon_0 $$ for all directional derivatives $D_u$
with respect to unit vectors  tangent to
$\partial \Omega_0 $ at $y$. Then, if the
constant $\epsilon_0$ is sufficiently small, one
can extend $x_t$ smoothly from $\partial
\Omega_0\times \Bbb R_t$ to $\{(y,t)\in \Bbb
R^n\times \Bbb R_t: |y|\leq \rho \}$ with the
following constraints\vskip.1in \noindent (a)
$||{\partial x_t \over
\partial y}(y,t)||<1.$ Here ${\partial x_t\over
\partial y}$ is the Jacobian matrix, and we are assuming that its matrix norm is less than one.
\vskip.1in\noindent (b) $x_t(y,t+1)=x_t(y,t)$ and
$\int_0^1 x_t(y,t)dt=0$. Note that $\int_0^1
x_t(y,t)dt=0$ for $y\in
\partial \Omega_0$. \vskip.1in \noindent (c) $|x_t(y,t)|<1$ for all $(y,t)$, and  $x_t(y,t)=0$ for $\rho^\prime \leq |y|\leq \rho$.
\vskip.1in
 Given (a)-(c), we define
$$\Psi(y,t)=y+\int_0^tx_t(y,s)ds.$$
Then $\Psi(y,t+1)=\Psi(y,t)$, and for $t\in
[0,1]$ the Jacobian matrix $\Psi_y$ satisfies
$$||\Psi_y(y,t)-I||\leq ||\int_0^t{\partial x_t\over \partial y}(y,s)ds||<1.$$
Therefore $\Psi(y,t)$ is a diffeomorphism of
$\Omega_0$ onto $\Omega_t$ with the properties
that ensure that $\Phi(y,t)=(\Psi(y,t),t)$ will
satisfy conditions (i)-(iv). \vskip.1in \noindent
{\bf 4. Asymptotically stationary motion.}
Stefanov considered the case of boundaries which
are stationary when $|t|$ is sufficiently large
in [St]. A generalization of this is the case
where the vector field $v(x,t)$ from \S 4.2
satisfies
$$\int_{\Bbb R}\sup_{\{|x|\leq \rho\}\cap \Omega(t)}||{\partial v\over \partial x}(x,t)||dt<\infty$$
Since the Jacobian of the mapping $F(t,y)$
satisfies
$${d\over dt}{\partial F\over \partial y}={\partial v\over \partial x}(F,t){\partial F\over
\partial y},$$
standard stability results (Theorem 1.1, Chpt. X
in [Ha]) imply that $||{\partial F\over
\partial y}(y,t)||$ is uniformly bounded. Thus, defining $\Phi(y,t)=(F(y,t),t)$ as in \S 4.2,
we have a mapping satisfying (i)-(iv).

The proof that the Jacobian, ${\partial F\over
\partial y}(y,t)$, is uniformly bounded is
particularly simple in this case: set
$\psi(t)=\sup_{\{|x|\leq \rho\}}||{\partial
v\over
\partial x}(x,t)||$, and let $w$ be one of the
columns in the Jacobian, $w={\partial F\over
\partial y_j}$. Then
$${1\over 2}{d\over dt}|w|^2=w\cdot {dw\over dt}=w\cdot {\partial v\over \partial x}(F,t)w\leq
\psi(t)|w|^2,$$ and for $t>0$ Gronwall's
inequality implies $|w(t)|\leq
|w(0)|\exp(\int_0^t \psi(s)ds)$. Similarly for
$t<0$ one has $|w(t)|\leq |w(0)|\exp(\int_t^0
\psi(s)ds)$.
 \vskip.2in\centerline{\bf References}\vskip.1in \noindent [CS1] Cooper, J. and
Strauss, W., Energy boundedness and decay of
waves reflecting off a moving obstacle, Indiana
U. Math. J., {\bf 25} (1976), 671-680.\vskip.1in
\noindent [CS2] Cooper, J. and Strauss, W., The
leading singularity of a wave reflected by a
moving body, J. Diff. Eq., {\bf 52}(1984),
175-203.\vskip.1in \noindent
 [CS3] Cooper, J. and Strauss, W., Abstract scattering theory for time-periodic systems
with applications to electromagnetism, Indiana U.
Math. J. {\bf 34}(1985), 33-83. \vskip.1in
\noindent [Ha] Hartman, P., {\it  Ordinary
Differential Equations}, Wiley\& Sons, New York
1964. \vskip.1in \noindent [Hi] Hirsch, M.,{\it
Differential Topology}, Springer-Verlag, New York
1976
 \vskip.1in \noindent[H\"o] H\"ormander, L., {\it
Analysis of Linear Partial Differential
Operators}, III, Springer-Verlag, New York 1985.
\vskip.1in\noindent [I] Ikawa, M., On the mixed
problem for hyperbolic equations of second order
with the Neumann boundary condition, Osaka Math.
J. {\bf 7}(1960),203-223. \vskip.1in\noindent
[In] Inoue, A., Sur $u_{tt}-\Delta u+u^3=f$ dans
un domaine noncylindrique, J. Math. Analysis and
App. {\bf 46}(1974), 777-819. \vskip.1in\noindent
[RZ] Robbiano, L. and Zuily, C., Uniqueness in
the Cauchy problem for operators with partially
holomorphic coefficients, Invent. Math {\bf
131}(1998), 493-539. \vskip.1in \noindent [St]
Stefanov, P., Inverse scattering problem for
moving obstacles, Math. Z. {\bf 207} (1991), 409
461-480. \vskip.1in\noindent [T] Tataru, D.,
Unique continuation for operators with partially
analytic coefficients, J. Math. Pures Appl. {\bf
78}(1999), 505-521.
 \end